# Evidence for the propagation of 2D pressure pulses in lipid monolayers near the phase transition


**J. Griesbauer[1,2], S. Bössinger[1,2], A. Wixforth[1], M.F. Schneider[2]**

*[1] University of Augsburg, Experimental Physics I, D-86159 Augsburg, Germany*

*[2] Boston University, Dept. of Mechanical Engineering, Boston-Massachusetts, USA*



**Abstract**

The existence and propagation of acoustic pressure pulses on lipid monolayers at the air/water-interfaces are directly observed by simple mechanical detection. The pulses are excited by small amounts of solvents added to the monolayer from the air phase. Employing a deliberate control of the lipid interface compressibility $\kappa$, we can show that the pulses propagate at velocities, which are precisely reflecting the nonlinear behavior of the interface. This is manifested by a pronounced minimum of the sound velocity in the monolayer phase transition regime, while ranging up to 1.5 m/s at high lateral pressures. Motivated by the ubiquitous presence of lipid interfaces in biology, we propose the demonstrated sound propagation as an efficient and fast way of communication and protein modulation along nerves, between cells and biological units being controlled by the physical state of the interfaces.




Lipid monolayers are often used as a model for biological systems. They also represent a simple and precise way to deliberately control the state of a soft interface. For example, a monomolecular film (monolayer) on water can be easily compressed on a film balance while the lateral pressure $\pi$ is monitored. This lateral pressure is connected to a very important thermodynamic property of the lipid film, its isothermal compressibility $\kappa_T$ being the negative inverse of the first derivative of the measured slope of $\frac{\partial \pi}{\partial A}$ at constant temperature.

While static properties, including phase transitions of 1[st] and 2[nd] order, have been studied intensively [1][2][3], there is only little literature on dynamic experiments. Experiments using harmonically oscillating barriers [4][5][6][7][8], capillary waves on water [9][10] or optically triggered molecules [11] were mainly used to analyze the complex dynamical compressibility $\kappa$ of different surfactants on water. Apart from this, also high frequency, electrical excitation of wave phenomena on lipid monolayers have been examined using a set of planar electrodes [12].

Here, we study the propagation of acoustic pulses along the lipid monolayer interface by directly observing the mechanical response of the excitable medium. In particular, our experiments prove the existence of propagating pressure pulse waves with amplitudes up to 0.3 mN/m. We demonstrate that the measured propagation velocities are controlled by the thermodynamic state of the monolayer reflecting the same nonlinearity $c(\pi)$ as the compressibility $\kappa(\pi)$. We present a simple linear hydrodynamic model which quantitatively describes our experimental findings and the measured velocities.

Lipid monolayers of 1,2-dipalmitoyl-*sn*-glycero-3-phosphocholine (DPPC) are spread from Chloroform to the air/water-interface of a film balance trough [12][13] (Fig. 1a). After *10* minutes of evaporation, the lateral pressure-area isotherm ($\pi$-A) is recorded by slowly ($\sim 2,5 Å^2/(min \cdot molecule)$) compressing the film, employing a moveable barrier. The trough is equipped with two pressure sensors (Whilhelmy plates), which can be read out very rapidly (*10000 Samples/second, 0,01 mN/m resolution*). These pressure sensors then are used to directly and mechanically read out the response of the arriving pressure pulses. The high sample rates allow for time resolved measurement of the response and for Fourier



transformation of the wave form. In order to exclude any spurious effects of water waves during the measurement of the two dimensional longitudinal pressure waves in the monolayer, an additional barrier is introduced in the trough (Fig. 1a). The compression pulses in the lipid layer are excited in a separate compartment by a sudden addition of a small amount of solvent (~ *3μl*) to the monolayer surface (Ethanol, Methanol, Chloroform, Pentane). Because of the additional barrier only waves which are able to travel over macroscopic distances and pass the small gap will cause the two pressure sensors to respond. To check if water waves are not affecting the sensors, references with the solvents added to pure water surfaces were recorded and indeed do not show any recognizable signals (see supplementary).

In Fig 1b, a typical result of a propagating pulse following an excitation is shown. The pulse was excited at *t~1s* by an Ethanol droplet. Shortly after excitation, a well defined pressure signal arrives at the Wilhelmy sensors at slightly different times. The time delay between the two pressure sensors and their given separation in space (*~14,5cm*), then allows to directly calculate the propagation velocity of the pressure pulse within the sensor compartment. To assure that the pressure pulse indeed travels the anticipated path, the additional barrier can be moved to the other side of the trough, opening a small gap near the excitation point (right). Indeed, the order in which the pressure sensors responded is exactly reversed. Varying the path length by changing the position of excitation, does not produce any resolvable differences in measured pulse heights, which implies non detectable attenuation of the pulses over the scale of the used trough (*20cm x 30cm*). Indeed, our theoretical description (see below) as well as the preceding experiments ([6][7][8]) suggest damping constants of *~1m⁻¹* rendering the excited monolayer pulses damping as technically immeasurable over the subjected propagation path (*~20cm*).

In Figure 2, the temporal response of the Wilhelmy plate for four different solvents in the liquid-expanded phase of the monolayer is shown (Fig. 2a). The pulse shapes appear not to be an inherent property of the film but to depend on the exciting solvent. While pulses excited from Pentane and Chloroform exhibit long trails (~20s), pulses of Ethanol and Methanol produce rather sharp pulses. This observation probably reflects the different residence times of the solvents in the monolayer, which in turn depends on the bulk/lipid partition coefficient of the solvent. Subsequent to the temporal pulse recording the spectrum of the excited pulses is analyzed by a Fourier Transformation. These spectra reveal that the highest observable frequencies (~ *1Hz*) depend only very little on the solvent.



We have recently shown ([12]), that high frequency electrical stimulation of a lipid film leads to wave phenomena with corresponding velocities around 100 m/s correctly predicted by

$$c_0 = \sqrt{\frac{1}{\rho_0 \kappa_S}} \qquad (1)$$

Here, $\rho_0$ is the film surface density and $\kappa_S$ is approximated by the isothermal lateral compressibility $\kappa_T$ directly extracted from the measured $\pi$-$A$-isotherm. In contradiction to equ. (1), our findings in this work as well as those of earlier studies [4][6][7][8][9][11] reveal that the observed propagation velocity of the pulses in the lipid film is very slow (~1m/s). One simple way to resolve the puzzle would be to account for viscous effects.

In this respect, the viscous force on a periodically ( $v_m = A e^{i(\omega t - k x)}$ ) excited lipid monolayer coupled to a water subphase is given by [6]

$$f \approx e^{i\frac{\pi}{4}} \sqrt{\eta_w \rho_w \omega} \, v_m \qquad (2)$$

Here, $\rho_w, \eta_w$ denote the density and viscosity of water [6][14]. Neglecting transversal components and assuming $\kappa_S$ to be frequency independent, we can introduced the viscous effects as an additional force into the Euler equation: $\rho_0 \dfrac{\partial v_m}{\partial t} = \dfrac{-\partial \Pi}{\partial x} - e^{i\frac{\pi}{4}} \sqrt{\eta_w \omega \rho_w} \, v_m$ . Using the continuity equation, one arrives at a wave equation for a monolayer being viscously coupled to the water underneath (see supplementary for details).

$$\frac{\partial^2 v_m}{\partial t^2} + \frac{1}{\rho_0} e^{i\frac{\pi}{4}} \sqrt{\eta_w \rho_w \omega} \frac{\partial v_m}{\partial t} - c_0^2 \frac{\partial^2 v_m}{\partial x^2} = 0 \qquad (3)$$

Using the ansatz $v_m = A e^{i(\omega t - k x)}$ , the dispersion relation for $\omega(k)$ can be extracted. For low frequencies $\omega$ the resulting attenuation $\beta$ and propagation velocity $c$ then turn out to be given by:



$$c = \frac{\omega}{\Re(k)} = \cos^{-1}(\frac{\pi}{8}) \sqrt{\frac{1}{\kappa_S} \sqrt{\frac{\omega}{\eta_w \rho_w}}} \qquad (4)$$

$$\beta = -\Im(k) = \sin(\frac{\pi}{8}) \sqrt{\kappa_S \sqrt{\eta_w \rho_w \omega^3}} \qquad (5)$$

As indicated by equ. (4), the velocity c is reduced with respect to $c_0 = \sqrt{\frac{1}{\rho_0 \kappa}}$ (1) by a factor

of order $\cos^{-1}(\frac{\pi}{8}) \sqrt{\rho_0 \sqrt{\frac{\omega}{\eta_w \rho_w}}}$, which, for $\omega \sim$ 1Hz is in the order of $\sim 10^{-3}$. Since only

insoluble lipid monolayers (DPPC) were investigated, further details like the diffusion of

lipids into the subphase have not been treated here, but have been discussed elsewhere in the

context of capillary wave theory [4][5].

In Fig. 3, we compare our experimental results to equ. (4) for all four solvents investigated.

The dotted lines represent the directly measured velocities, whereas the solid lines have been

calculated using $\omega \sim 1Hz$ from the measured pulse form (Fig. 2) and the independently

evaluated isothermal compressibility $\kappa_T = \frac{-1}{A} \left( \frac{\partial A}{\partial \pi} \right)_T$. The velocities are plotted as a

function of the applied lateral pressure $\pi$. As can be seen we find very good agreement for the

entire range of compressibilities and for all four solvents. Both linear and nonlinear (region of

maximum in $\kappa$) regimes including the characteristic dip in the velocity profile at the

maximum of the compressibility are clearly resolved in model and experiment. The velocity

decreases from $c=1.5 \; m/s$ in the liquid condensed phase to $c=0.4 \; m/s$ in the liquid expanded

phase with a minimum of $c=0.2 \; m/s$ in the transition regime, demonstrating the close relation

between thermodynamic state and propagation velocity.

For the evaluations above the assumption was made, that $\kappa_T \approx \kappa_S$, which is not true in

general and may explain the small differences of measurement and theoretical expectations in

Fig. 2. Equ. (4) and (5) only hold exactly when one uses the adiabatic compressibility $\kappa_S$.

Hence, the real dynamic or adiabatic compressibility $\kappa_S$ as a function of area or lateral

pressure may be extracted using Equ. (4) and measurements of the propagation velocity.

Comparing this $\kappa_S$ to $\kappa_T$ we find, that according to thermodynamic expectations [14], $\kappa_S < \kappa_T$



over the entire pressure spectrum. As indicated before ([5][15]), the maximum of $\kappa_S$ is less pronounced than that of $\kappa_T$ ( see supplementary).

Finally, we studied the degree of excitability in the different thermodynamic states of the interface. Therefore, the detected peak height of the propagating pressure pulse was analyzed for different lateral pressures and is plotted in Fig. 4. Close to the maximum of the compressibility the detected pulse signal becomes very weak and is barely detectable. This directly proves that the thermodynamic state of the interface, which is controlled by the lipid monolayer, controls not only propagation speed but also its "strength".

In summary, we have given direct experimental proof that adiabatic longitudinal pulses can be mechanically/chemically excited and directly detected in lipid monolayers. Assuming an additional viscous drag force being caused by the interaction between lipid film and subphase (water underneath) we could show that the velocity and the amplitude of the propagating pulses both depend predictably on the thermodynamic state of the interface via its compressibility. Realizing the ubiquity of soft interfaces in biological systems, our experiments not only support the idea of sound as the foundation of nerve pulses [16][17][18], but suggest such pulses or 2D wave propagation in general as a very efficient way for intra and intercellular communication. In addition membrane proteins are of course exposed to the lateral conditions (mechanical, electrical, thermal etc.) of the membrane. If a protein's lateral motion is part of its function, pressure pulses as described here, will therefore certainly modulate this function. This may be of essential importance for mechanosensitive channels or pores [19][20][21][22]. Furthermore and in the authors opinion even more important, the pressure pulse is capable to change the state of the membrane interface. In static experiments, it has been shown that such a change not only regulates the activity of membrane bound enzymes [23][24][25], but is also capable to even induce "channel-like" fluctuations in the absence of proteins [26][27].

Our simple model and the equations derived can of course not account for these additional anticipated effect. To clarify this exciting question, at least two fundamental aspects need still to be evaluated: i) nonlinear response contributions near the maximum compressibility and ii) the thermodynamic coupling of *all* the observables (area, charge, pH, heat, etc.). It will be interesting to investigate to what extend pulse propagation, channels and catalytic activity can be simply explained thermodynamically.

**Figure 1 a)** Film balance used for Monolayer-Pulse-Analysis. The trough is equipped with two pressure sensors and an additional barrier, which separates the excitation site from the detection compartment. **b)** Time course of the sensor readouts for a pulse travelling from sensor 2 to sensor 1 on a DPPC-Monolayer (*24°C*). Before the pulse signal arrives at sensor 2, it had to propagate through the small gap between additional barrier and the trough walls. To assure that the pressure pulse indeed travels the anticipated path, the additional barrier has been moved to the left side of the trough opening the small gap near the excitation point (right). In fact, the order in which the pressure sensors responded was exactly reversed.

**Figure 2** Response of pressure sensor 1 for the excitation with four different solvents in the liquid expanded phase of a DPPC-Monolayer (*24°C*). Obviously the measured pulse shape depends on the solvent. Pentane and Chloroform, for example, remain longer in the monolayer (*~10s*) before complete evaporation. Correspondingly, the recorded pulse extends over a longer period of time. A FFT frequency analysis of the pulses suggest dominant frequencies up to *~1Hz*.

**Figure 3** Propagation velocities of the pulses, determined by the run time differences between sensor 2 and 1 (distance *~14,5cm*) for different lateral pressures of the DPPC-Monolayer (*24°C*). Additionally, the model according to equ. (2) and the isothermal compressibility $\kappa_T$ are plotted into the corresponding graphs for an average pulse frequency of *~1Hz*. Both curves coincide very well, even the typical $(\kappa_T)^{-1}$ minima in the phase transition regime of the DPPC-Monolayer (*24°C*) are well reproduced.

**Figure 4** Measured pulse heights as a function of lateral pressure in the lipid film (DPPC, *24°C*). Like the propagation velocity, a distinct minimum evolves at the compressibility maximum indicating the phase transition regime. Hence, both pulse velocities and heights depend on the phase state of the DPPC-Monolayer.



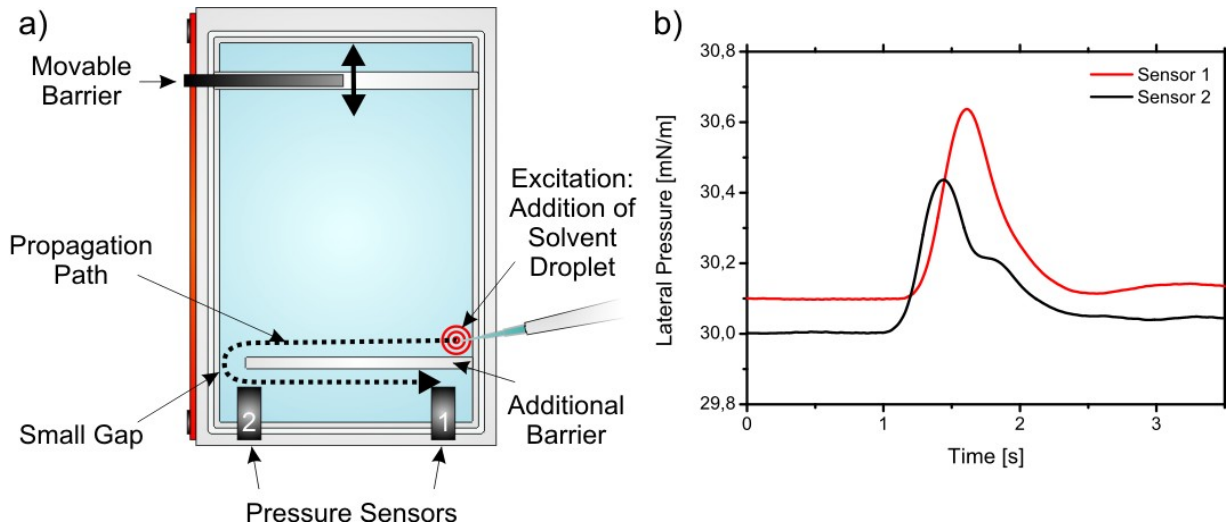

**Figure 1**

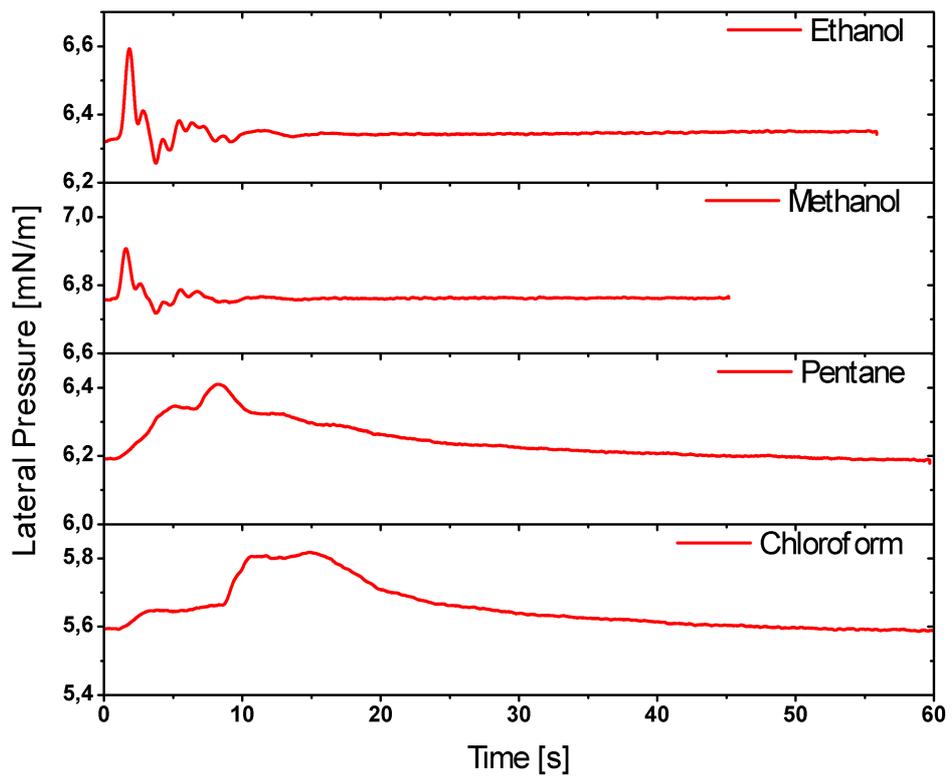

**Figure 2**



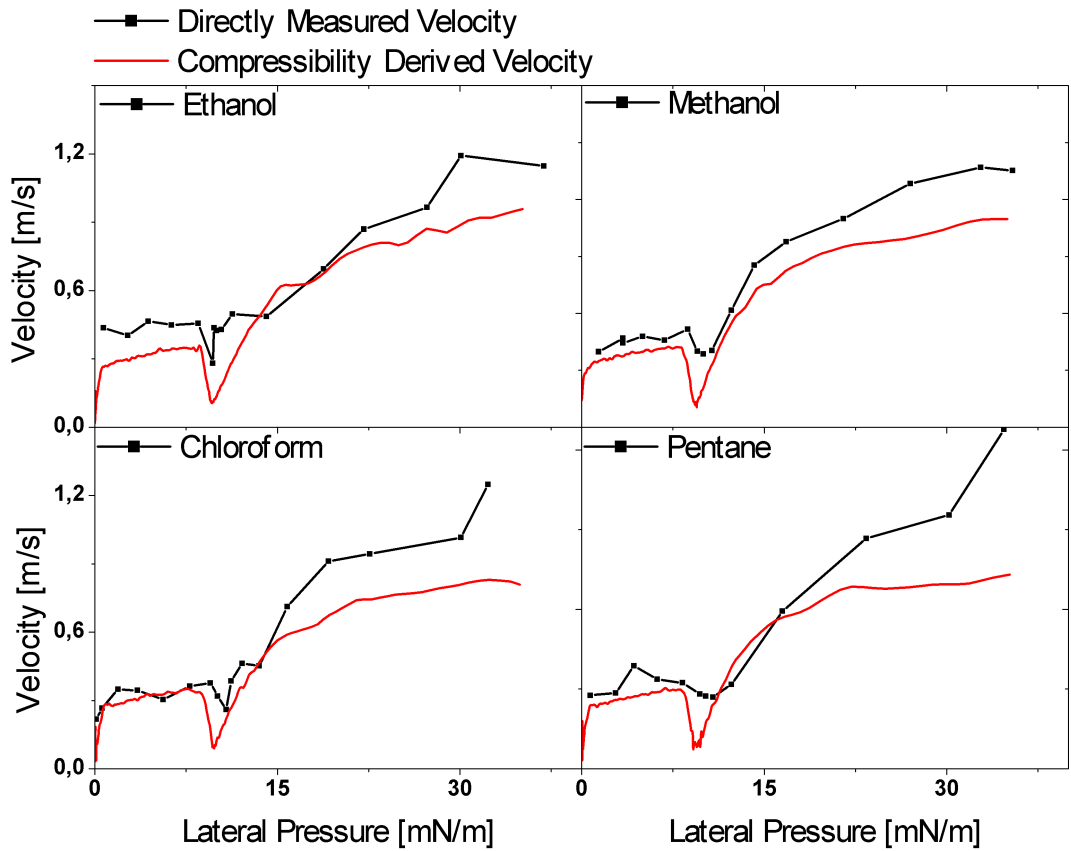

**Figure 3**

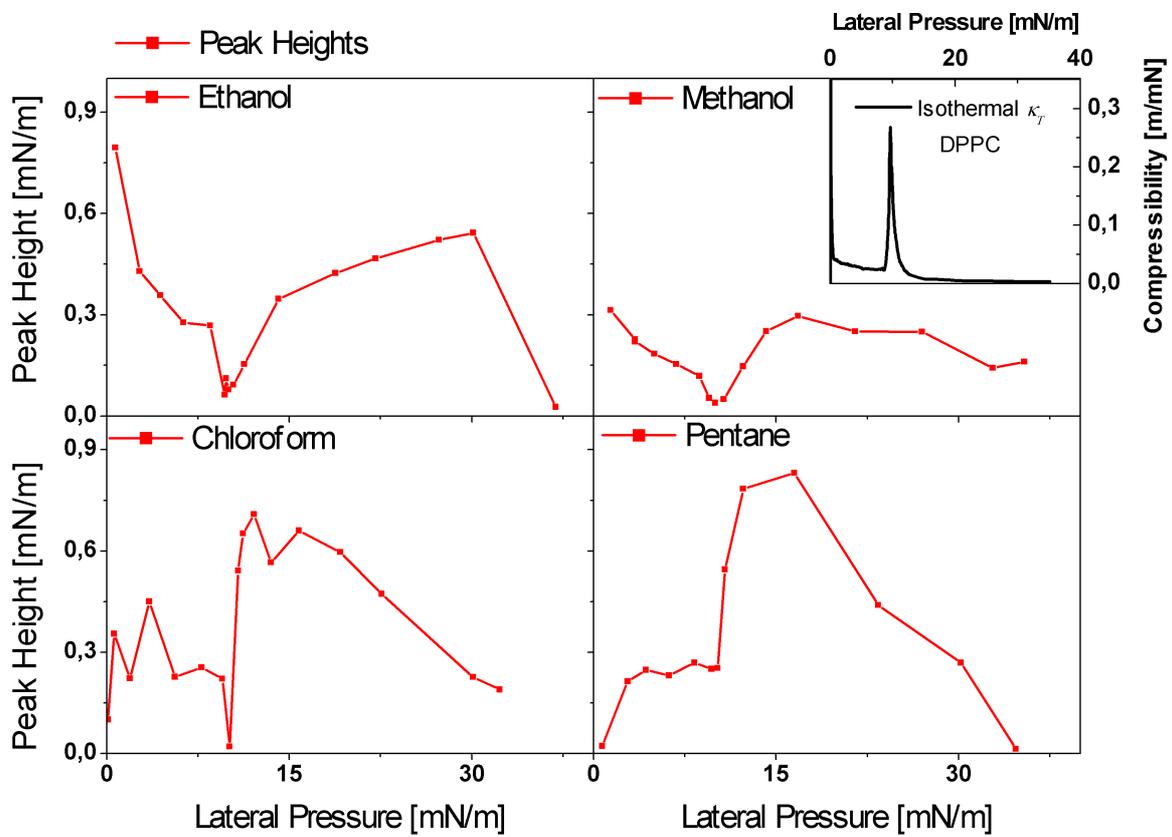

**Figure 4**